\documentclass{ceurart}

\usepackage{tabularx}
\usepackage{booktabs}
\usepackage{amsmath}
\usepackage{array}
\usepackage{xspace}
\usepackage{graphicx}
\usepackage{rotating}

\newcommand{\VQA}{Video Question Answering\xspace}
\newcommand{\OR}{\mathrm{OR}}

\begin{document}

\copyrightyear{2026}
\copyrightclause{Copyright for this paper by its authors. Use permitted under
Creative Commons License Attribution 4.0 International (CC BY 4.0).}

\conference{XAI.it 2026: 6th Workshop on Explainable AI, Generative and
Agentic Systems: Trust, Transparency, and Human Oversight, hosted at
AIxIA 2026, October 2026, Perugia, Italy}

\title{Following Delegation Traces to Study Human Oversight in Interaction with Multimodal AI}

\author[1]{Agnese Chiatti}[
  email=agnese.chiatti@polimi.it
]
\cormark[1]
\author[1]{Matteo Matteucci}
\author[1]{Viola Schiaffonati}

\address[1]{AI and Robotics Lab (AIRLab), Politecnico di Milano, Milan, Italy}
\cortext[1]{Corresponding author.}

\begin{abstract}
As multimodal AI systems are becoming our everyday thought partners, the shift towards studying how users transfer control to these systems becomes especially urgent. In this paper, we report an overview of findings from an online study with $n=56$ AI-literate participants and 392 Video Question Answering trials. After inspecting each multimodal AI response, participants rated task-specific trust and selected one of three intended policies for an analogous future task: reject AI assistance, use it with verification, or delegate fully to the AI. Within participants, post-interaction trust strongly predicted the user's delegation choices ($\OR=14.97$, 95\% CI
$[7.14,31.37]$). Trust and delegation nevertheless remained distinct. Verification was the dominant policy and often co-occurred with high trust. Delegation also increased with task order after adjustment for momentary and participant-level trust ($\OR=1.22$, 95\% CI $[1.02,1.45]$), indicating an experience-related component that self-reported trust ratings alone could not capture. These findings motivate a shift from intended delegation to a non-intrusive observation of how delegation is enacted in collaborative human-AI tasks. We therefore conclude by proposing enacted delegation traces as a methodological framework for future studies of multimodal and agentic AI use.
\end{abstract}

\begin{keywords}
human-AI delegation \sep
human oversight \sep
explainable AI evaluation \sep
multimodal AI \sep
repeated interaction \sep
behavioural measurement
\end{keywords}

\maketitle

\section{Introduction}
\label{sec:introduction}

A typical day in the life of a 2026 knowledge worker may look like this. The day begins in an online meeting, where an AI assistant listens quietly in the background, turning the discussion into notes. Later, while drafting a technical report, they ask another assistant to revise a paragraph and keep only the most useful suggestions. A scientific claim looks useful, but not quite safe to trust, so they verify it against external sources. They also need a chart to accompany the text, so they ask another AI assistant to write the code for generating it, before adding their final touches.  

AI systems are no longer only tools for retrieving information. They are becoming our everyday thought partners \cite{collins2024building}. We ask them to interpret
material, draft outputs, advise on various topics and domains, and, in emerging agentic systems,
even initiate actions on our behalf \cite{acharya2025agentic}. 

Human agency and oversight are central to European frameworks for Trustworthy AI and are explicitly reflected in the AI Act \cite{eu2019ethics,eu2024aiact}. Yet translating
these principles into interaction-level evaluation remains an open challenge: technical properties such as transparency, robustness, and accountability
\cite{chander2024trustworthy,novelli2024accountability} do not by themselves reveal how users retain or transfer control while working with an AI system. We therefore ask: \textit{how can human oversight be made observable in practice while AI carries out part or all of a task?}

We propose delegation as a complementary lens for studying human oversight as it is enacted in interaction with multimodal AI systems. Delegation concerns which parts of a 
task users hand over to an AI system, how they inspect or verify its outputs, 
and when they intervene to change or reclaim control. In this view, a system may satisfy 
technical and regulatory requirements for trustworthiness while still eliciting poorly calibrated 
delegation. For instance, a user may trust an AI system while still deciding to verify its output or completely avoid full delegation for tasks they perceive as personally meaningful or risky. The same user, nonetheless, may delegate to a less trusted system when time,
fatigue, or lack of expertise makes independent work costly. For this reason, we distinguish: 
\textit{trustworthiness}, as the properties of the system and its deployment; 
\textit{trust}, as the user's situated perception of those properties 
\cite{mayer1995}; and \textit{delegation}, as the observable 
transfer of control over task execution and decision authority between the human and the AI system \cite{lee2004}. This perspective is also relevant to the Explainable AI (XAI) field: an explanation or inspectable representation can support oversight not only by changing trust, but by shaping how users delegate to the AI the system.

Existing human-AI research has extensively examined trust and users' responses to AI recommendations, while comparatively less work studies when users transfer control to AI and how much oversight they retain \cite{gomez2025,mehrotra2024,gor2026}. This gap is particularly relevant for multimodal systems, where existing evaluation still focuses predominantly on model performance and user preferences \cite{lee2024vhelm,zhang2024multitrust,lu2024wildvision,chou2025visionarena} rather than on how control is negotiated across repeated interactions.

%Delegation provides a complementary perspective to trust. Building on classical views of trust and reliance in automation \cite{mayer1995,lee2004,hoff2015}, in this paper we make a distinction between \textbf{trust} as \textit{the perceived trustworthiness of an AI system from the user’s perspective}, and \textbf{delegation} as \textit{the observable behavioural choice to rely on the system for part (or all) of a task}. 

The contribution of this paper is twofold. First, we provide a task-level account of how trust, repeated experience, and the availability of an inspectable graph representation for the model answers relate to 
intended delegation among AI-literate users, 
whom we selected as a deliberate first population for studying these dynamics 
during repeated multimodal interaction (Section 3 and 4). Second, we introduce \textit{enacted delegation traces} as a general framework 
for studying how responsibility is redistributed during collaborative tasks 
across user populations and application contexts (Sections 5 and 6). We 
formalise these traces through the scope of control transfer, the oversight actions performed after delegation, and the recovery of control. We then 
outline how these elements may be observed under varying time pressure, 
perceived AI competence, and task stakes. Experimental results reported in this paper measure intended delegation only; the enacted-delegation framework introduced in Sections 5–6 is a methodological proposal motivated by these findings that has not yet been implemented or validated.
Our proposed framing positions delegation as 
a measurable object for evaluating the quality of oversight supported by 
multimodal and agentic AI systems.

In the paper, we use a focused empirical result to motivate a methodological
position. 
Video Question Answering provides a controlled multimodal setting in which users must judge whether an AI system has correctly represented objects, actions, and temporal relations before deciding how much control to retain. From the standpoint of XAI, it also enables a
comparison between a textual answer and an editable spatio-temporal graph that exposes part of the represented video situation for inspection. Thus, we designed a repeated \VQA study, where participants rated trust after each response and selected whether they would reject AI assistance, use it with retained control, or delegate fully on an analogous task. Higher-than-usual trust strongly predicted stronger intended delegation, while retained control remained common even at
high trust. Exploratory findings further suggest that graph-supported presentation was associated with stronger delegation. Together, these findings show that delegation captures how users allocate oversight across repeated interaction in ways that trust ratings alone leave only partly visible.

\section{Background and Related Work}
\label{sec:background}

Trust has long been treated as a central construct for explaining how people use automated systems. In classic accounts, it describes a willingness to accept vulnerability to another actor \cite{mayer1995}, while appropriate reliance concerns whether users rely on automation when it is capable and withhold reliance when it is not \cite{lee2004}. This distinction is important for human-AI interaction because trust remains a subjective judgement, whereas reliance and delegation become visible in how users allocate control. Delegation is especially relevant for systems that do more than provide advice. Yet empirical work still tends to study the users' post-interaction reactions to AI recommendations more often than the conditions under which users are prepared to transfer control. Gomez et al., for instance, found delegation in only a small fraction of the human-AI interaction sequences reviewed in their survey \cite{gomez2025}. Recent studies have begun to operationalise delegation more directly, either through willingness to delegate a repeated perceptual task \cite{mehrotra2024}, or through behavioural traces of whether teams allow an AI partner to answer on their behalf \cite{gor2026}. These works motivate measuring delegation as a construct related to trust, but analytically distinct from it.

Delegation also connects to established work on human-automation function allocation. Parasuraman et al. distinguish types and levels of automation across information acquisition, analysis, decision selection, and action implementation \cite{parasuraman1997}. Pritchett et al. frame function allocation as assigning work across human and automated agents, highlighting potential mismatches between responsibility and authority \cite{pritchett2014measuring}. More recent human-AI research \cite{adam2024navigating} shows that whether delegation is initiated by a system or user can affect system acceptance. From an Explainable AI (XAI) perspective, explanations can support appropriate reliance by helping users distinguish when AI advice should or should not be followed \cite{schemmer2023appropriate,mehrotra2024}. More generally, making the evidence or intermediate representation behind an answer inspectable can support oversight by giving users a basis for checking and revising AI contributions
\cite{cosseddu2025explainable}. 

The distinction between reported trust and delegation is particularly underexplored in multimodal AI, compared to language-based AI models and tasks. Research on vision-language models has largely focused on model-centric trustworthiness metrics, including hallucination, robustness, and benchmark performance, while fewer studies examine how users interpret repeated multimodal outputs and translate these evaluations into delegation decisions \cite{chiatti2025mapping}. Benchmarks such as VHELM and MultiTrust broaden evaluation beyond accuracy by considering several dimensions of multimodal model behaviour \cite{lee2024vhelm,zhang2024multitrust}. Public agent arenas \cite{lu2024wildvision,chou2025visionarena} move closer to situated use by collecting open-ended interactions and preference judgements. However, user preferences do not show how trust evolves across repeated tasks, nor how users retain and transfer their judgment or control over a task. A co-evolutionary view of human-AI interaction further suggests that these choices should be studied across repeated encounters, as both user strategies and system capabilities change through use \cite{pedreschi2025humanaicoevolution}.

Our enacted-delegation perspective builds on these ongoing efforts by focusing on how control transfer during task execution, answer inspection, and recovery of control unfold behaviourally during interaction. We begin exploring Video Question Answering tasks, where users must assess whether a system has correctly interpreted objects, actions, and temporal relations in a video. The STAR benchmark is well suited to this setting because its questions depend on situated reasoning over structured events \cite{wu2021}. In such tasks, a fluent textual answer can conceal the intermediate representation on which the answer depends. Graph-based presentations offer one way to expose this intermediate layer by making entities and relations inspectable \cite{mitra2024compositional,lusha2025graph}. In our study, we treat graphs as an interface-level presentation model rather than as a guarantee of correctness. It externalises part of the represented video situation, gives users a concrete object to inspect and edit, and allows us to test the effects of a more interactive interface on self-reported trust and delegation choices.

\section{An Empirical Study with AI-literate Users}
\label{sec:study}

\subsection{Participants and task design}

We conducted an online study involving participants from Politecnico di Milano aged 19-50 years old ($M=24.43$, $SD=4.79$). 45 (80.3\%) studied Computer
Science and Engineering, and the remaining 19.7\% came from other engineering majors. 43 were enrolled in a Master's, PhD or postgraduate study program. In line with the gender distribution in the broader population, the sample included 42 men (75.0\%) and 14 women (25.0\%). We deliberately began with a relatively young and technically experienced 
population familiar with AI, allowing us to examine delegation among users 
already accustomed to evaluating AI outputs before extending the proposed 
approach to populations with different levels of experience. 

We recruited participants through internal student mailing lists, student-representative channels, the AIRLab LinkedIn account, and campus leaflets with QR codes. Of 101 respondents who completed the pre-screening and consent questionnaire, 87 accessed the study platform and 56 completed a full study session (i.e., exactly 7 trials). Participation was voluntary and uncompensated, and participants were informed that the study would require approximately 15 minutes. One participant completed three sessions; to avoid disproportionate weighting, we retained only their first completed session for inference. The final sample therefore comprised \(N=56\) participants and 392 analysed task observations.

\begin{figure*}[t]
    \centering
    \caption{Overview of the experimental groups and conditions.}
    \label{fig:experimental-groups}
    \includegraphics[
        width=0.75\textwidth,
        trim={41bp 73bp 58bp 74bp},
        clip
    ]{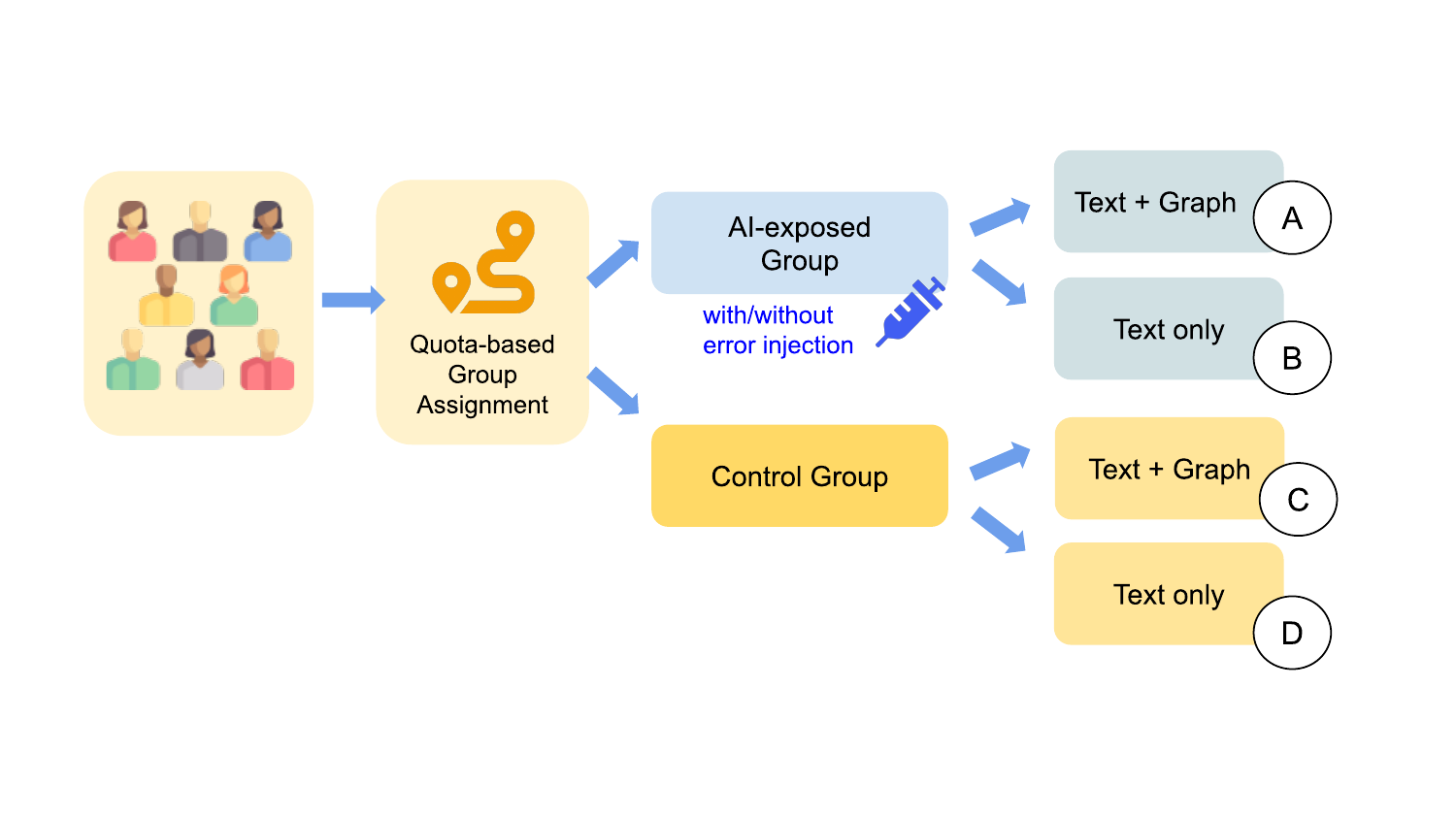}
\end{figure*}

Each retained participant completed one guided practice task followed by exactly seven analytic \VQA trials derived from the STAR benchmark \cite{wu2021}. The analytic task set comprised the same seven selected STAR sequence items across participants; these required connecting actions occurring at different points in an everyday video and determining their temporal order. System responses were generated in advance with Gemini 2.5 Flash so that participants assigned to the same experimental condition encountered stable and comparable stimuli. \\
We restricted the STAR material to temporal sequence questions because preliminary experiments \cite{lusha2025graph} found temporal and predictive questions among the most challenging for the Gemini 2.5 Flash. Since predictive questions in STAR include relevant answer evidence only in the ground-truth graphs, but not in the accompanying video clips, we limited our analysis to sampled temporal questions to enable a fair comparison between text-only and graph-supported variants.  \\
In each trial, participants
first watched the video and answered the multiple-choice question. They then
inspected a response presented as coming from a generic multimodal AI, without knowing any details about the specific model used for generating answers. We tested four different website variants enabling controlled experimental conditions, summarised in Figure~\ref{fig:experimental-groups}. Depending on
the group assignment, users were exposed on either text-only answers or text accompanied by an editable
spatio-temporal graph. Moreover, two samples served as control group: i.e., users thought they were evaluating AI-generated answers when in fact they were presented with ground truth data. To control for the presence of errors in the AI-generated responses, a later participant cohort encountered deliberately
erroneous responses. The present paper does not delve into the analysis of condition effects.
Here we simply provide an overview of the pooled effects among trust, intended delegation, and task
order, with condition indicators retained as covariates. No a priori or post-hoc power analyses were conducted: small and uneven between-condition samples are treated as exploratory rather than precisely estimated intervention effects.

\begin{table}[t]
\centering
\caption{Experimental design and first-survey inferential sample. 
Each participant contributed seven analytic trials. Error injection and 
randomisation of video order were introduced only in the later cohort.}
\label{tab:design}
\small
\begin{tabular}{llclrr}
\hline
Variant & Cohort & Graph & Response source & $n$ & Trials \\
\hline
A & Early & yes & AI-generated          & 5  & 35 \\
A & Late  & yes & AI-generated + errors & 10 & 70 \\
B & Early & no  & AI-generated          & 11 & 77 \\
B & Late  & no  & AI-generated + errors & 3  & 21 \\
C & Early & yes & ground truth          & 6  & 42 \\
C & Late  & yes & ground truth          & 7  & 49 \\
D & Early & no  & ground truth          & 9  & 63 \\
D & Late  & no  & ground truth          & 5  & 35 \\
\hline
  & Early total & & & 31 & 217 \\
  & Late total  & & & 25 & 175 \\
\hline
  & Overall     & & & 56 & 392 \\
\hline
\end{tabular}
\end{table}

Participants were assigned to interface variants using an adaptive quota-balancing procedure rather than simple random assignment. On first access, the platform assigned each participant to the least-filled variant; ties were resolved according to the priority A, C, B, D, with additional preference for graph-supported variants when these were under-represented. Once assigned, participants retained the same variant throughout the study. The study used the same seven analytic video stimuli across participants. Video order was fixed in the earlier cohort and randomised in the later cohort.

\begin{sidewaysfigure}[p]
    \centering
    \scriptsize

    \begin{minipage}[t]{0.49\textwidth}
        \centering
        \textbf{(a) Initial answer and pre-response trust}\\[0.2em]
        \includegraphics[width=\linewidth]{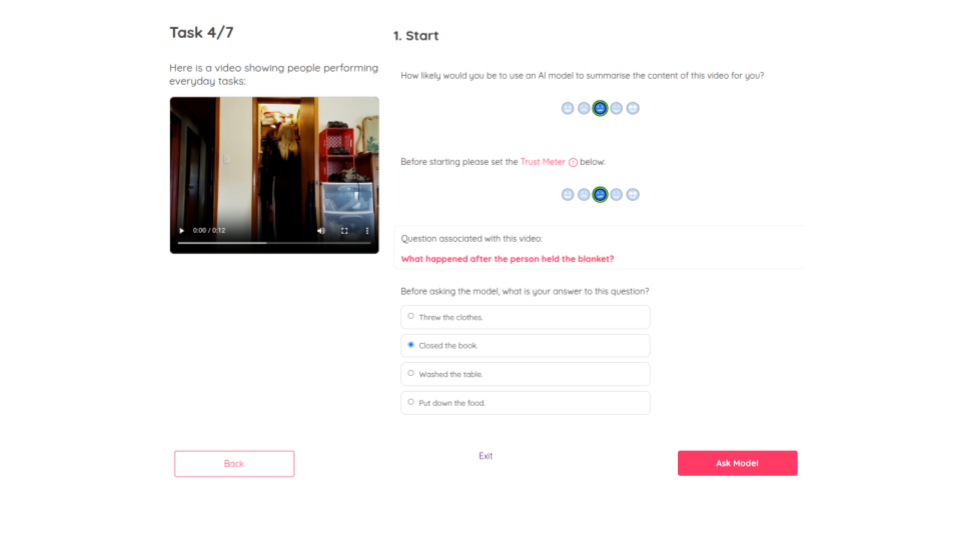}
    \end{minipage}
    \hfill
    \begin{minipage}[t]{0.49\textwidth}
        \centering
        \textbf{(b) Model-response inspection}\\[0.2em]
        \includegraphics[width=\linewidth]{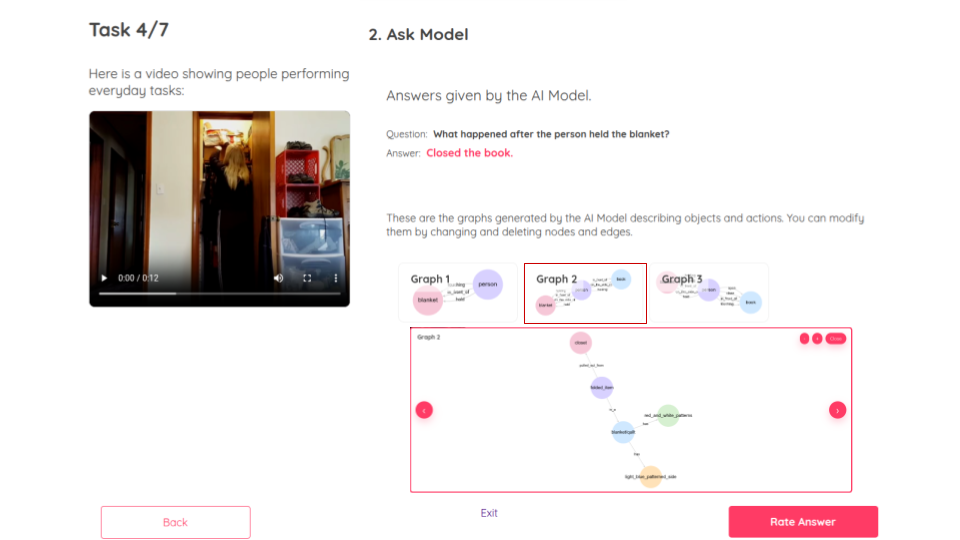}
    \end{minipage}

    \vspace{1em}

    \begin{minipage}[t]{0.49\textwidth}
        \centering
        \textbf{(c) Response evaluation}\\[0.2em]
        \includegraphics[width=\linewidth]{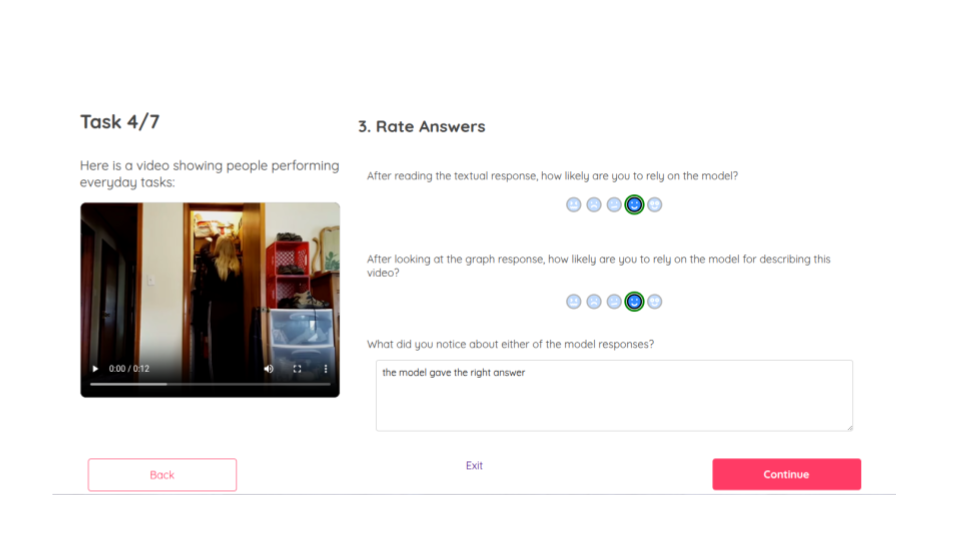}
    \end{minipage}
    \hfill
    \begin{minipage}[t]{0.49\textwidth}
        \centering
        \textbf{(d) Post-response trust and delegation choice}\\[0.2em]
        \includegraphics[width=\linewidth]{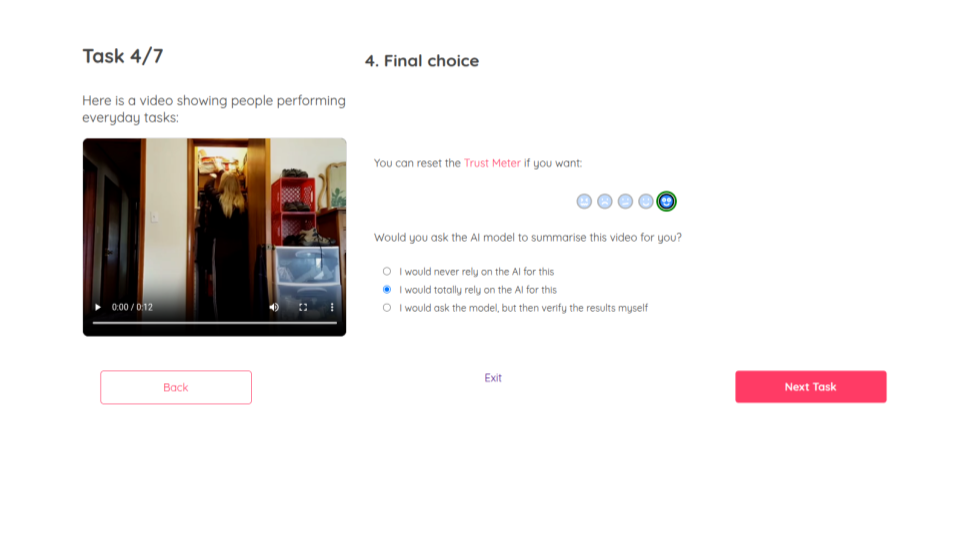}
    \end{minipage}

    \caption{Four-stage trial interface. The video remained visible throughout the sequence, ending with post-response trust and delegation choice.}
    \label{fig:gui}
\end{sidewaysfigure}

At the beginning and end of each trial (Figure\ref{fig:gui}a), participants rated current trust in the system through
a five-point pictorial meter, stored from 0 (very low) to 4 (very high). 

Then, after inspecting (Figure\ref{fig:gui}b) and evaluating the AI responses (Figure\ref{fig:gui}c),
they answered whether they would ask the AI to summarise an analogous video by
choosing one of three ordered policies:
\begin{equation}
  \text{reject} < \text{use with verification} < \text{fully delegate}.
  \label{eq:delegation-order}
\end{equation}

The middle option specified that the user would check the AI output
independently. Because the choice concerned a future analogous task rather than
an actual delegation performed in the trial, it is interpreted as \emph{intended}
delegation. 

The "final choice" page (Figure~\ref{fig:gui}d) was conceived to capture users' post-response trust rating immediately before their intended-delegation decision. This design also allowed us to compare pre- and post-interaction trust within participants across trials, accounting for the possibility that trust varied between videos and questions.

The browser-based platform used an HTML5/JavaScript frontend with a PHP/MySQL backend. Interactive graphs were rendered with Cytoscape.js. The online survey platform we developed is showcased in Figure~\ref{fig:gui} and publicly available at \url{https://www.refine-webapp.polimi.it/}. The code for reproducing the results reported in this paper can be found at \url{https://github.com/achiatti/refine-study-preliminary}. We have also released to the public the fully anonymised survey data on Zenodo \cite{chiatti2026perceivedtrust}.

The study received institutional ethical and data-protection
clearance before recruitment. Participants provided their informed consent and were free to withdraw their participation at any time. 

\subsection{Within-person analysis}

The repeated observations were analysed with a cumulative-logit mixed model and
a participant-specific random intercept. Post-response trust was separated into
within-person and between-person components. For participant $i$ on task $t$,

\begin{equation}
  \mathrm{WithinTrust}_{it}=
  \mathrm{PostTrust}_{it}-\overline{\mathrm{PostTrust}}_i.
\end{equation}

The pooled model predicted the ordered delegation choice from within-person
trust, the participant's mean trust, task order, recruitment cohort, graph
presentation, and error-exposure assignment. This model allowed us 
to separate two levels of variation: between-person differences captured by each participant’s mean post-response trust, and within-person fluctuations captured by trial-level trust relative to that participant’s own mean. We used within-person trust as focal predictor; recruitment cohort, graph presentation, and error exposure were included as adjustment covariates, while task order was retained as an exploratory predictor (Table \ref{tab:ordinal_model}).
Therefore, the focal
coefficient asks whether the same participant selected a stronger
policy on trials in which their trust was higher than their personal average. This within-person estimate is important because it reduces the risk of
confusing stable individual differences with task-level changes in trust and
delegation. We used likelihood-ratio testing to assess whether the focal trust term improved model fit. \\
We also report the association with task order as an exploratory result, since the study 
was not designed to determine whether changes across the session reflected learning, 
fatigue, habituation, or growing familiarity with the AI support. Because task order and item identity were confounded in the earlier cohort, we additionally re-estimated the pooled cumulative-logit model including the seven video items as categorical fixed effects. The participant random intercept and all other predictors were unchanged. \\ 
Since the present 
paper overviews the study mainly to motivate the transition towards enacted 
delegation, we provide only a compact analysis of graph presentation alongside 
the pooled relationships among trust, intended delegation, and task order. 

\begin{table}[t]
\centering
\caption{Complete fixed-effect estimates from the pooled cumulative-logit
mixed model predicting intended delegation ($N=56$, 392 trials).}
\label{tab:ordinal_model}
\small
\begin{tabular}{lrrr}
\hline
Predictor & $\beta$ (SE) & OR [95\% CI] & $p$ \\
\hline
Within-person trust
    & 2.71 (0.38) & 14.97 [7.14, 31.37] & $<.001$ \\
Mean participant trust
    & 4.39 (0.54) & 80.42 [28.01, 230.86] & $<.001$ \\
Later cohort
    & 0.79 (0.60) & 2.21 [0.68, 7.19] & .187 \\
Graph presentation
    & 4.39 (0.69) & 80.95 [20.91, 313.33] & $<.001$ \\
Error exposure
    & -1.76 (0.67) & 0.17 [0.05, 0.63] & .008 \\
Task order
    & 0.20 (0.09) & 1.22 [1.02, 1.45] & .030 \\
\hline
\end{tabular}
\vspace{0.5em}

\begin{minipage}{0.95\linewidth}
\footnotesize
\textit{Note.} Thresholds: $\alpha_1=11.44$, $\alpha_2=19.90$.
Participant random-intercept variance = 19.08 (SD = 4.37).
\end{minipage}
\end{table}

\section{Selected Findings}
\label{sec:results}

\subsection{Delegation with retained control remained common at high trust}

Across the 392 trials, participants chose verification in 230 cases (58.7\%),
rejected AI assistance in 114 (29.1\%), and selected full delegation in 48
(12.2\%). The joint distribution in Figure~\ref{fig:trust-delegation-heatmap} shows that higher
trust was generally associated with stronger intended delegation, while leaving
substantial variation within the same trust level (Spearman's $\rho=.522$).

\begin{figure}[t]
\centering
\caption{
Joint distribution of post-response trust and intended delegation across 392 trials. Cell entries are task counts.
}
\label{fig:trust-delegation-heatmap}
\setlength{\tabcolsep}{2pt}
\setlength{\extrarowheight}{0pt}
\renewcommand{\arraystretch}{1.08}

\begin{tabular}{
    @{}
    >{\raggedright\arraybackslash}p{0.27\linewidth}
    *{5}{>{\centering\arraybackslash}m{0.105\linewidth}}
    @{}
}
\toprule
&
\multicolumn{5}{c}{\textbf{Post-response trust}}
\\
\cmidrule(lr){2-6}

\textbf{Delegation choice}
& \textbf{0}
& \textbf{1}
& \textbf{2}
& \textbf{3}
& \textbf{4}
\\
\midrule

Reject AI assistance
& \cellcolor{blue!5}2
& \cellcolor{blue!28}30
& \cellcolor{blue!50}56
& \cellcolor{blue!24}26
& \cellcolor{blue!0}0
\\

Use with verification
& \cellcolor{blue!0}0
& \cellcolor{blue!18}20
& \cellcolor{blue!58}64
& \cellcolor{blue!85}\textcolor{white}{109}
& \cellcolor{blue!34}37
\\

Fully delegate
& \cellcolor{blue!0}0
& \cellcolor{blue!3}1
& \cellcolor{blue!5}2
& \cellcolor{blue!17}18
& \cellcolor{blue!25}27
\\

\bottomrule
\end{tabular}
\end{figure}

\begin{figure}[!t]
    \centering
    \caption{Four-zone classification of post-response trust and delegation. Points are trial-level observation and is jittered to make overlapping observations 
visible. The off-diagonal zones capture high trust with retained 
oversight and full delegation under low trust.}
\label{fig:trust-delegation-zones}
    \includegraphics[width=0.75\linewidth, trim={0.5cm 0.5cm 1cm 1.5cm}, clip]
{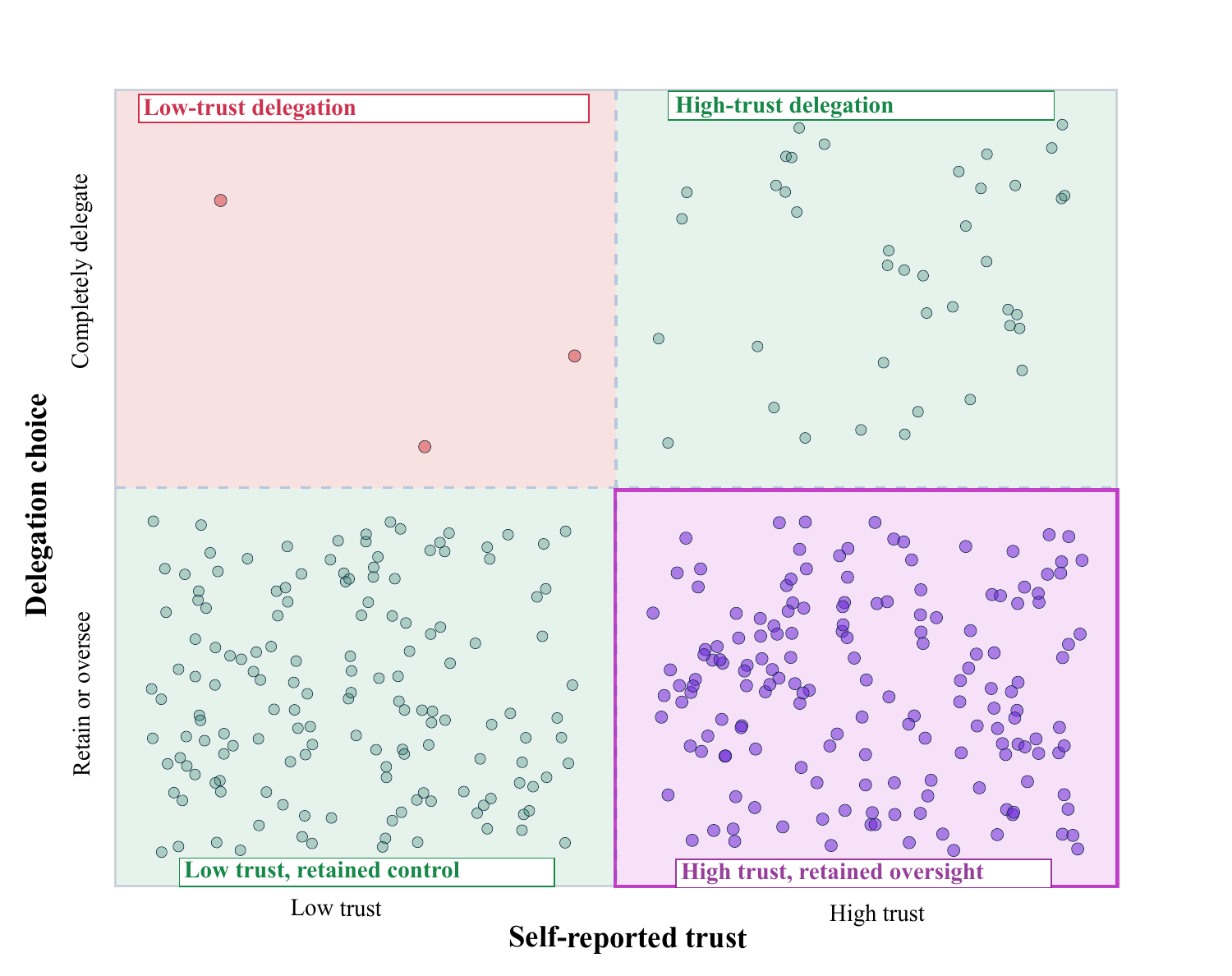}
\end{figure}

The divergence was asymmetric. With ratings 3 and 4 treated conservatively as
high trust, 172 trials combined high trust with retained human control through
rejection or verification, as is visible in the four-quadrant analysis of Figure \ref{fig:trust-delegation-zones}. This represents 79.3\% of all high-trust trials.
Only three trials combined low trust (0-2 ratings) with full delegation. A favourable
evaluation therefore often coexisted with oversight, whereas delegating while evaluating the system poorly was rare. This pattern corroborates monitoring the interplay of self-reported trust and delegation as non-overlapping computational objects of interaction.

\subsection{Task-specific trust predicted stronger intended delegation}

A one-point increase in post-response trust relative to a participant's own mean
was associated with approximately fifteen times higher odds of selecting a
stronger delegation category ($\OR=14.97$, 95\% CI $[7.14,31.37]$). Removing the
within-person trust term substantially worsened model fit
($\mathrm{LR}\,\chi^2(1)=74.979$, $p<.001$). This estimate indicates that, for the same person, a task
on which momentary trust was higher than usual was also a task on which the person
reported greater willingness to transfer control. This finding makes post-interaction trust behaviourally consequential in terms of the selected policy for intended delegation.

\subsection{Inspectable graph presentation and delegation}
All 28 participants assigned to a graph-supported condition interacted with the 
editable spatio-temporal representation, and 16 (57\%) modified or deleted at 
least one node or edge. This confirms that the representation functioned as an interactive inspection mechanism rather than merely as additional visual information. In the exploratory pooled model, graph presentation was 
associated with stronger intended delegation after adjustment for trust, task 
order, cohort, and error exposure. However, the estimate was highly uncertain because the graph-supported groups were small ($n=15$ and $n=13$) and unevenly distributed across cohorts.

Graph presentation did not reliably predict trust-delegation mismatch 
($\mathrm{OR}=1.33$, 95\% CI $[0.55, 3.23]$; 
LR $\chi^{2}(1)=0.354$, $p=.552$), and the cohort-specific estimates differed 
in direction. The graph may therefore have increased participants' willingness to delegate without improving trust-delegation alignment. Based on these results, exposure to inspectable graphs should not be treated as evidence of better-calibrated oversight, under the operationalisation of mismatch we used in these analyses (i.e., the four-zone analysis of Figure~\ref{fig:trust-delegation-zones}). 

\subsection{Intended delegation showed a task-order association}

Each successive task was associated with higher odds of selecting a stronger
intended delegation category ($\OR=1.22$, 95\% CI $[1.02,1.45]$, $p=.030$),
after adjustment for momentary trust, participant-level trust, and the study
conditions. This association was essentially unchanged when video identity was additionally included as a categorical fixed effect ($\OR = 1.24$, 95\% CI $[1.00, 1.53]$, $p=.048$), suggesting that the pooled task-order association was not solely attributable to differences among the seven items.

Trust, on the other hand, did not display a corresponding session-wide trend.
The task-order association persisted after accounting for item identity, though we interpret this result as exploratory rather than evidence that repeated interaction caused greater delegation. Participants may have learned the task,
become more comfortable with AI assistance, or may have experienced fatigue as the task progressed. The current design cannot distinguish these mechanisms, also because
the selected policy had no effect on the next trial. However, this evidence motivates extending the study to a more collaborative setup where these mediating factors are tracked more systematically.

\section{Position: From Intended to Enacted Delegation}
\label{sec:position}

The intended delegation item used in the study we just overviewed was task-specific, temporally close to the trust rating, and preserved verification (i.e., delegation with retained control) as an explicit oversight policy. These properties made it useful for showing that trust and delegation can be measured together during repeated multimodal AI interaction. At the same time, the measure remained separated from the collaborative process itself. Participants did not hand over part of the current task, their choice did not alter the workflow, and no performance or effort consequences followed from the selected policy. The verification option also compressed behaviours that may differ substantially in practice, such as briefly checking a salient video segment, reconstructing the answer independently, or correcting an intermediate representation before accepting the AI contribution, for instance.

We therefore propose moving from intended delegation to \emph{enacted delegation traces}. We define enacted delegation traces as the observable redistribution of task responsibility between a human and an AI system during an ongoing activity. In this view, delegation is the behavioural process through which users decide what the AI should do, what they will inspect, and when they will reclaim control. The relevant unit of analysis is no longer only the final decision to accept or reject assistance, but the trace of delegations and interventions that occurs while the task is being completed.

For a task episode \(t\), we represent enacted delegation as a compact trace:
\begin{equation}
\mathcal{D}_{t} = \langle s_{t}, o_{t}, r_{t} \rangle ,
\label{eq:delegation-trace}
\end{equation}
Here, the three components denote dimensions to be operationalised from interaction logs rather than fixed scalar measures: \(s_t\) captures the scope of the control transfer, \(o_t\) the oversight performed after delegating, and \(r_t\) recovery of control and its consequences.
Scope concerns how much of the task is assigned to the AI system and how central that component is to the final outcome. Oversight concerns the checking work performed by the user, including inspection of evidence, revision of AI content, and independent attempts to verify the output. Recovery concerns whether the user detects a problem, interrupts or reverses the AI contribution, and restores human control when needed. 

Crucially, these dimensions should be analysed jointly. Indeed, broad delegation followed by timely recovery may reflect more effective oversight than narrow delegation accompanied by no substantive inspection. Moreover, these elements should be interpreted together with the conditions under which delegation takes place. Indeed, in collaborative AI use, the same transfer of control may have different meanings depending on time pressure, perceived AI competence, and the perceived stakes of the task. For the proposed study, we focus on three non-exhaustive contextual factors and associate each episode with a  contextual vector:
\begin{equation}
\mathbf{x}_{t} = \langle \tau_{t}, \gamma_{t}, \pi_{t} \rangle ,
\label{eq:contextual-vector}
\end{equation}
where \(\tau_{t}\) denotes time pressure, \(\gamma_{t}\) denotes perceived AI competence, and \(\pi_{t}\) denotes perceived risk or incentive. These factors are important because users may delegate to save time, to compensate for limited expertise, or because the consequences of an error are perceived as tolerable. Conversely, they may retain oversight even under high trust when the task feels consequential or when accountability remains personally salient.

This formalisation treats \textbf{delegation as a computational object of human-AI interaction}. The task interface we envision to implement this formalisation can reconstruct \(\mathcal{D}_{t}\) from ordinary workflow events. Thus, interaction events such as assigning a component to the AI model, inspecting its output, editing the AI contribution, rejecting generated content, or returning to manual control for completing a task will all have to be recorded by the proposed platform. Specifically, these events should be logged at the adequate level of granularity to reconstruct the moment-by-moment evolution of \(\mathcal{D}_{t}\). This in-situ reconstruction is important because delegation is rarely a single decision made once at the beginning or end of a task. A participant may initially assign the temporal reconstruction to the AI, then inspect only one critical video segment, accept part of the generated summary, and later reclaim control after noticing a mistake in the model answer. Collapsing this sequence into a final label such as ``verified'' or ``delegated'' would miss the trajectory through which oversight was actually performed. Momentary tracking therefore allows delegation to be studied as a dynamic process.

We distinguish this delegation trajectory from the contextual conditions under which it unfolds. This distinction can be expressed schematically as
\begin{equation}
\mathcal{D}_{t} \sim
p_{\theta}\!\left(\mathcal{D}_{t}\mid\mathbf{x}_{t},\mathbf{z}_{t}\right),
\end{equation}
where $\theta$ denotes model parameters and $\mathbf{z}_{t}$ collects additional
task-, system-, and participant-level covariates, such as interface condition,
AI reliability, task order, and prior AI familiarity. 
This notation keeps delegation analytically distinct from the factors that shape it. Time pressure, for instance, does not constitute delegation by itself, but it may increase the probability of delegation or reduce the degree of inspection post-interaction. Similarly, perceived AI competence may make users more willing to assign central components to the AI, while higher perceived stakes may increase verification, especially after a suspected error. The proposed expression is a
conceptual template for future operationalisation rather than a model estimated in the present study.

In this framing, self-reported trust remains central as a subjective measure, but it should be positioned as complementary to the delegation trace. Sparse trust probes can be collected before the task, at predetermined checkpoints during the interaction, or after controlled failures. This additional recording of trust ratings would reveal regions where self-reported trust and enacted delegation overlap or, more interestingly, diverge. However, we argue that these self-reported ratings should remain sparser and intentional to avoid disrupting the unobtrusive observation of how delegation is embedded in real use, as proposed in the following section.  

\section{Proposed Study Design for Enacted Delegation}
\label{sec:enacted-delegation}

\textbf{Design considerations.} The next methodological step is to observe how delegation is enacted while a human and an AI system jointly produce an outcome. As defined in Section~\ref{sec:position}, we introduce \emph{enacted delegation traces} to make observable the transfer of control between human and AI during an ongoing activity. The study we envision should therefore seamlessly integrate delegation as part of the task, rather than an additional question asked after the task has already ended.

A promising direction is to extend the Video Question Answering setup into a more realistic and gamified learning scenario. Participants could watch short recorded lessons or tutorials and answer a sequence of questions under varying time constraints. Sample questions may require, for instance, locating a specific moment in the video, reconstructing the order of explained steps, or applying concepts explained in the video to a new case. The number of questions could be feasible in the baseline condition and deliberately difficult in groups where we aim to test the effect of time pressure. The interface must also allow for hybrid delegation scenarios: participants could ask the AI to locate relevant video segments, draft an answer, or produce a short explanation, while still being responsible for the final submitted response.

This task directly extends the empirical study reported in this paper and can 
initially be applied to the same population of AI-literate university students. 
This population provides a useful starting point for refining the measures and 
interaction traces with users already familiar with AI-supported work. At the same time, it is closer to everyday AI use in educational settings, where students increasingly rely on AI systems to summarise lectures and prepare for assessments. However, the framework is general and can be applied to the study of broader populations, e.g., by integrating more informational and general-purpose videos, or workplace training material. 

The interface should instantiate delegation through ordinary workflow actions mimicking real-world use. Participants should be able to work manually, assign a component to the AI, or take control back after inspecting the AI contribution. Once AI content is available, the participant can inspect the relevant video evidence, revise the answer, or accept the contribution into the final response. The system records 
the events needed to reconstruct control transfer and oversight as these unfold. Sparse trust probes can still be collected at specific checkpoints, so that subjective trust remains available without interrupting every micro-decision.

In any case, the interface design should be preceded by a participatory phase. Practical workshops and focus groups can help us identify the situations in which different users already delegate to AI, the moments in which they would like improved and increased oversight of the system. Moreover, there may be key factors shaping delegation in real use beyond time pressure, perceived capability and risk that we may have not anticipated before a focused discussion with users. Users may also respond to boredom, uncertainty about assessment criteria, social norms around acceptable AI use, or effort-related factors, to name just a few. Thus, this participatory phase is instrumental in informing the final task scenarios and control items implemented in the envisaged interface.

\textbf{Modelled conditions.} An important requirement for conducting the envisaged study is ensuring that specific experimental controls are in place for factors shaping enacted delegation. One way to approach this requirement is to adopt an approach similar to the experimental setup presented in this paper, where participant groups are assigned to different control conditions while seemingly completing the same task. The first condition to test concerns the AI output presentation mode. Depending on the task, the interface could expose AI outputs structured differently: e.g., from a plain textual answer up until a richer interactive workspace where users can view, edit or reject intermediate outputs. This configuration would extend the graph-presentation manipulation used in the preliminary study, while shifting the role of the representation from an explanation shown after the answer to an object through which the user can supervise the collaboration. 

A second ablation concerns the behaviour of the AI assistant. One dimension for controlling the AI behaviour is the injection of incorrect and fabricated answers, similar to the error-injected cohort in our preliminary study. Another dimension is to expose users to different levels of proactive suggestion by the AI, to simulate agentic scenarios. Lastly, another interesting but still underexplored dimension concerns the comparison between a static AI system and adapting AI systems that co-evolve with users.  

Around these interface and AI-support conditions, the study should explicitly vary task context. Time pressure can be manipulated, for example, by changing the number of questions that must be completed within the available time, making full manual completion difficult in some conditions. Perceived AI competence can be shaped through a pre-task calibration phase in which participants see the assistant's prior performance on similar tasks. Task stakes can be assessed by introducing reward/penalty schemes, e.g., where participants earn points for correct answers and lose a small number of points when they do not provide supporting evidence alongside the answers. These incentives to make the user's choices more consequential and engaging ought to be introduced while making sure to keep the interaction between the user and the AI low-risk and non-harmful from an ethical standpoint.

The proposed design is different from work on cooperative question answering where delegation is observed mainly as the decision to let the AI answer on behalf of the team \cite{gor2026}. Here, delegation unfolds within a multimodal workflow in which users transfer control over specific parts of the task, inspect or revise AI contributions, and reclaim control. It also extends work on task delegability and instance-level delegation \cite{lubars2019delegability,spitzer2025human} by treating contextual factors as part of the experimental situation rather than as only post-hoc explanations for a final choice. The resulting delegation traces would make it possible to study how human oversight evolves while the task is still in progress.

\section{Conclusion}
\label{sec:conclusion}

This paper started from an empirical study that deliberately focused on 
AI-literate users as an initial population for examining how post-interaction 
trust relates to intended delegation in repeated Video Question Answering tasks. In this context, the choice to verify the AI outputs remained common even when self-declared trust was high. This pattern suggests that trust is behaviourally consequential, but demands a more systematic study of how human oversight dynamics are embedded in everyday use. The graph comparison adds an XAI-specific observation. Inspectable and editable representations were actively used and were 
exploratorily associated with stronger intended delegation, while showing no 
reliable association with trust-delegation mismatch. Future evaluations should 
therefore examine whether graph inspection helps users identify incorrect 
relations, revise the represented evidence, and retain control when errors are 
detected.

We thus proposed a shift toward enacted delegation, providing a methodological bridge between subjective trust and human oversight. In this definition, we framed delegation as a traceable redistribution of responsibility across a task, allowing researchers to observe how much work is handed over, to what extent AI outputs are inspected, and whether control can be recovered after failure. This perspective is especially relevant for multimodal and agentic AI systems, where assistance increasingly involves interpreting heterogeneous evidence and acting across several steps of a workflow.

Empirically, we showed that intended delegation captures aspects of repeated human-AI interaction that cannot be reduced to trust ratings. Methodologically, we proposed a framework to be operationalised and validated in future studies of delegation as it is enacted in collaborative multimodal tasks, while controlling for time pressure, perceived AI competence, and task stakes. Although the empirical study began with AI-literate participants, the proposed 
framework is intended to support comparative studies across populations with 
different levels of AI familiarity and domain expertise. 

Our broader aim is to encourage discussion on evaluating multimodal, explainable, and agentic AI systems in terms of the quality of oversight they support. Enacted delegation offers one way to make this quality observable by turning the small, distributed decisions through which users retain or transfer control in everyday AI use into a measurable object of human-centred interaction research.

\begin{acknowledgments}
The authors would like to acknowledge the funding received to support this work, through the Italian MUR SOE "Young Researchers" initiative (project ID SOE\_20240000048). 
\end{acknowledgments}

\section*{Declaration on Generative AI Use}
During
the preparation of this work, the authors used ChatGPT 5.5 to help with grammar and spelling checks, \LaTeX{} formatting, minor paraphrasing and rewording. After using these tools or services, the authors reviewed, verified,
and edited the content as needed and take full responsibility for the
content of this publication.

\bibliography{references}

\end{document}